# Measurement of a solid-state triple point at the metal-insulator transition in VO$_2$

Jae Hyung Park[1], Jim M. Coy[1], T. Serkan Kasirga[1], Chunming Huang[1], Zaiyao Fei[1], Scott Hunter[1], and David H. Cobden[1*]

[1]Department of Physics, University of Washington, Seattle WA 98195, USA

*Corresponding author: cobden@uw.edu

**First-order phase transitions in solids are notoriously challenging to study. The combination of change in unit cell shape, long range of elastic distortion, and flow of latent heat leads to large energy barriers resulting in domain structure, hysteresis, and cracking. The situation is still worse near a triple point where more than two phases are involved. The famous metal-insulator transition (MIT) in vanadium dioxide[1], a popular candidate for ultrafast optical and electrical switching applications, is a case in point. Even though VO$_2$ is one of the simplest strongly correlated materials, experimental difficulties posed by the first-order nature of the MIT as well as the involvement of at least two competing insulating phases have led to persistent controversy about its nature[1-4]. Here, we show that studying single-crystal VO$_2$ nanobeams[5-16] in a purpose-built nanomechanical strain apparatus allows investigation of this prototypical phase transition with unprecedented control and precision. Our results include the striking finding that the triple point of the metallic and two insulating phases is at the transition temperature, $T_{tr} = T_c$, which we determine to be 65.0 ± 0.1 °C. The findings have profound implications for the mechanism of the MIT in VO$_2$, but in addition they demonstrate the importance of such an approach for mastering phase transitions in many other strongly correlated materials, such as manganites[17] and iron-based superconductors[18].**

The MIT in VO$_2$ is accompanied by a large and rapid change in the conductivity and optical properties, with potential uses in switching and sensing. VO$_2$ has recently received renewed attention as a convenient strongly correlated material for the application of new ultrafast[19-21] and microscopy[22,23] techniques, ionic gating[24], and improved computational approaches[3,4]. However, the problems associated with bulk or film samples that consist of a complex of multiple solid phases and domains under highly nonuniform strain, as well as compositional variations such as oxygen vacancies[25] and hydrogen doping[26], make it almost impossible to disentangle the underlying parameters on which rigorous understanding can be built. The experiments described here eliminate these problems, allowing unprecedented control of the MIT and accurate determination of the underlying phase stability diagram of pure VO$_2$ for the first time.

Fig. 1a illustrates the structures of the phases involved in the MIT. In every phase there are two interpenetrating sets of parallel chains of vanadium atoms each surrounded by six oxygen atoms forming a distorted octahedron (the oxygen atoms are not shown). In the high-temperature metallic (rutile, R) phase all the chains are straight and periodic, whereas in the low-temperature insulating (monoclinic, M1) phase every chain is dimerized. There are also two other known insulating phases: monoclinic M2, in which only one set of chains is dimerized; and triclinic T, which is intermediate between M1 and M2. The existence of both M1 and M2, with similar dielectric yet different magnetic properties, provides constraints on the theory of the MIT; for example, it rules out a purely Peierls-type mechanism[2]. In the older literature the MIT is taken to occur between R and M1, although recent studies[8-10,23] have shown that M2 domains occur in most VO$_2$ samples near the MIT, raising the question of its role in the transition.

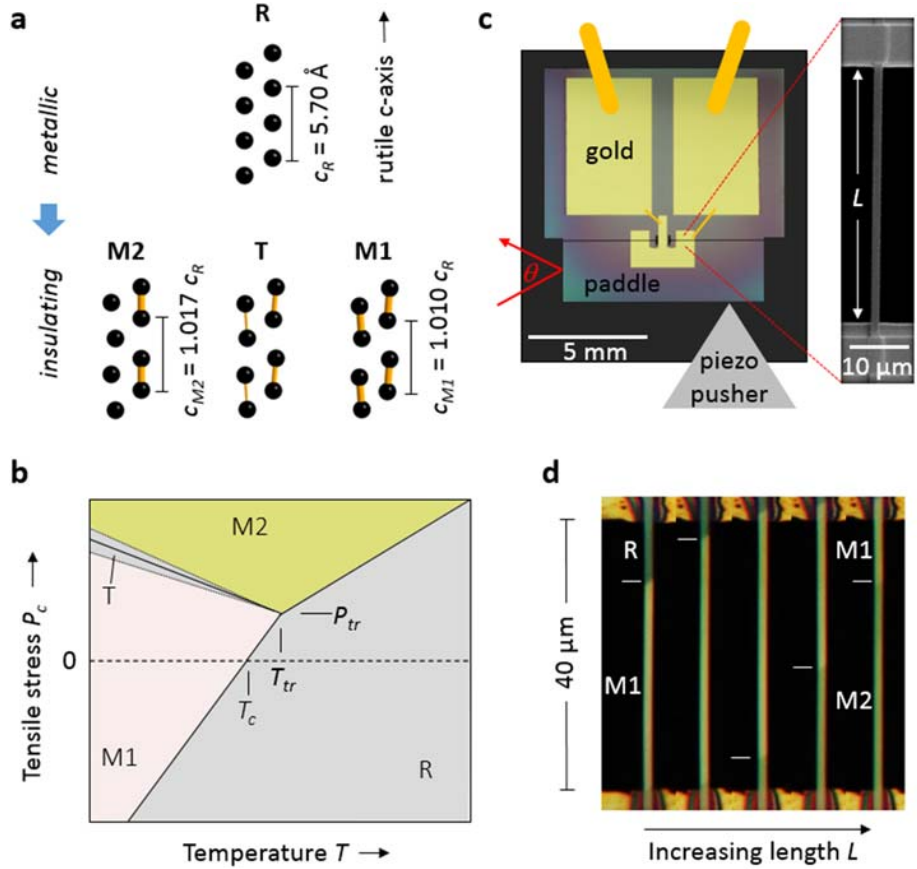

**Figure 1 | Control of the metal-insulator transition in VO$_2$ using uniaxial stress. a**. Arrangement of vanadium ions in the phases involved in the MIT, showing their different vanadium chain periods and dimerization (yellow). **b**. Expected layout of the stress-temperature phase diagram near the MIT, with the transition temperature $T_c$ at zero stress. **c**. Experimental geometry, showing an electron micrograph (right) of a VO$_2$ nanobeam suspended across a slot of width $L$ in a silicon chip (left, optical micrograph) whose width is controlled by pushing on the paddle and measured via deflection of a laser beam. The yellow lines signify gold wire bonds. **d**. Series of optical images showing movement of the R-M1 and M1-M2 interfaces as $L$ is increased in roughly 100 nm steps at 64 °C (device P7, 40 μm gap).

The largest difference in unit cell shape between R, M1 and M2 is along the pseudo-rutile *c*-axis (the vanadium chain axis), with $c_R = 5.700$ Å, $c_{M1} = 5.755$ Å, and $c_{M2} = 5.797$ Å, as indicated in Fig. 1a. Compressive strain along this axis in an epitaxial film can lower the transition to room temperature[15,25], and thus applying uniaxial tensile stress $P_c$ along it can be used to control the transition[13,15]. A stability diagram in the $P_c - T$ plane (with all other stress components zero) is expected to have the layout indicated in Fig. 1b. A shaded region indicates where the T phase occurs[7,27]. The effect of $P_c$ on the phase stability (Fig. 1b) resembles that of stress along the [110]$_R$ axis[27] and of doping[28] by Cr. Rough ideas of the locations of the three phase boundaries have been obtained by modeling bent nanobeams[16]. The triple point $(T_{tr}, P_{tr})$ has not been located, although M1 and M2 are known to be very close in free energy near the transition[27]. $P_{tr}$ is normally taken to be positive, implying that a perfect unstrained crystal exhibits a direct transition from M1 to R at $T_c$. We find that this is not in fact the case and, remarkably, $T_{tr}$ is identical to $T_c$ to within ±0.05 °C, or one part in $10^4$ in absolute temperature. We further determine $T_c$ to be 65.0 ± 0.1 °C. In addition we present evidence that in the neighborhood of $T_c$ the M1 phase can distort continuously under tension into the metastable T phase. These discoveries have deep implications for the physics of the MIT, for the interpretation of many measurements on VO$_2$ crystals and films, and for mastering the transition with a view to applications.

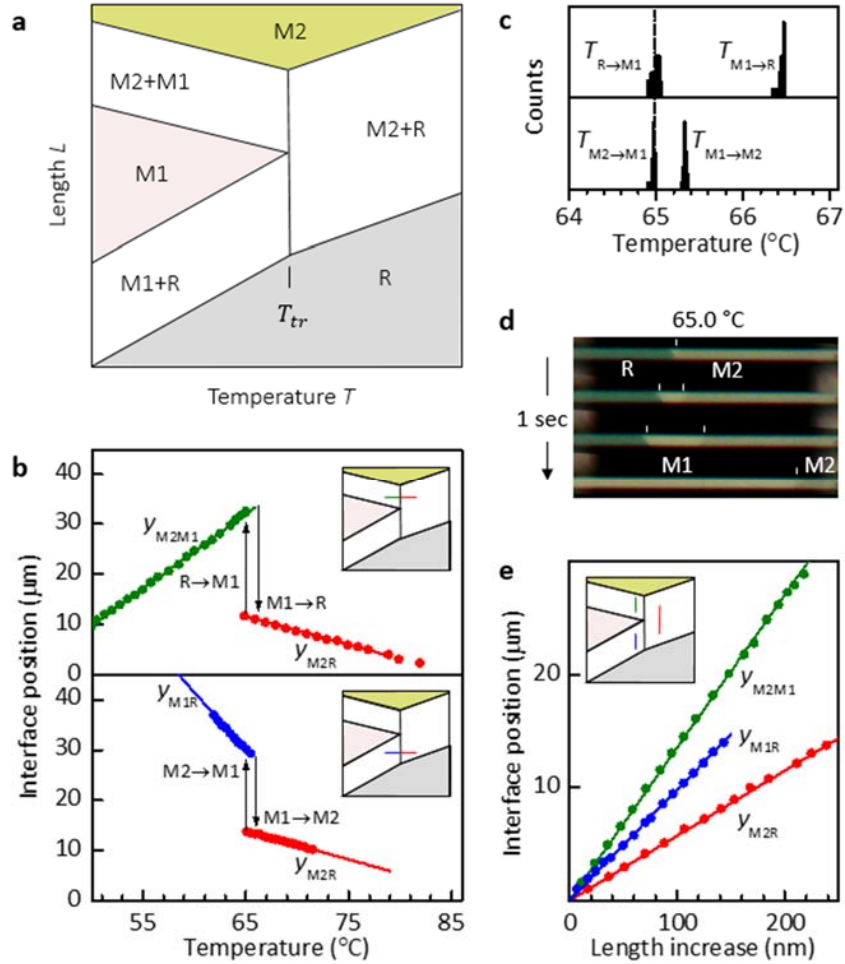

**Figure 2 | Temperature and length dependence in coexistence**. **a**, Expected configuration of a nanobeam as a function of $T$ and $L$. **b**, Variation of interface positions with $T$ at fixed $L$ corresponding to moving along the lines in the insets (upper: device P11, 40 μm gap; lower: P9, 20 μm). Each interface type is indicated by a color. c, Histograms of temperatures at which reconfigurations occur for 20 cycles sweeping at 0.1 °C/minute (P14, 40 m). d, Sequence of images during reconfiguration from M2+R to M2+M1 in a nanobeam at the triple point, 65.0 °C (P8B, 20 μm). **e,** Variation of interface positions with $L$ at fixed $T$, corresponding to moving along the vertical lines in the inset (P14). The fractional differences in lattice constants, $\alpha_{ij}$, are the inverse slopes of these lines.

Our investigations of the MIT rest on the ability to precisely control the length of a suspended single-domain nanobeam and thereby to apply pure uniaxial stress along it, a situation that cannot be achieved in larger crystals because of domain structure. The elements of the experiment are illustrated in Fig. 1c (see Methods). A VO$_2$ nanobeam is fixed, in some cases with electrical contacts, across a micromachined slot in a silicon chip whose width $L$ can be varied with nanometer precision. We perform measurements only when the nanobeams are straight, so the maximum compressive stress is limited by buckling. By varying $L$ and $T$, the three phases R, M1 and M2 can be induced and can be differentiated by reflection contrast using linearly polarized light[10], as illustrated in Fig. 1d, as well as by Raman spectroscopy[14] and electrical resistance measurements. Linearly polarized light also reveals twinning[11], allowing us to select devices where twinning is absent.

According to the phase diagram in Fig. 1b the state of the nanobeam as a function of $L$ and $T$ should include regions of two-phase coexistence as sketched in Fig. 2a. We find that indeed the suspended part of the nanobeam can be brought into coexistence between any pair of the three phases. The position of the interface changes smoothly and reproducibly with both $L$ and $T$ in between sudden reconfigurations. For the case of M2+R coexistence we define the interface position $y_{M2R}$ as the shift relative to an initial position such that it increases as R converts to M2. We define $y_{M2M1}$ and $y_{M1R}$ similarly.

The MIT in VO$_2$ is usually studied as a function of $T$, without paying close attention to strain or to interconversion between M1 and M2. In undoped samples it is seen in the range 65 – 68 ° C, with a hysteresis of several degrees, and the value of $T_c$ is not known more precisely than this. In our experiments on nanobeams, as $T$ is varied at fixed $L$ we see the behavior shown in Fig. 2b, which can be understood with reference to the color-coded lines in the inset phase diagrams. If we start in M2+M1 coexistence (upper panel, green) and increase $T$, the interface position $y_{M2M1}$ first moves smoothly as the stress required for phase equilibrium changes[13]. Then at a temperature $T_{M1 \to R}$ there is a sudden reconfiguration to M2+R coexistence (red) after which the interface position $y_{M2R}$ moves smoothly again. On cooling, the reverse reconfiguration occurs at temperature $T_{R \to M1}$. Starting instead at a smaller length, in M1+R coexistence (lower panel, blue), a jump to M2+R coexistence (again red) occurs at $T_{M1 \to M2}$, while the reverse occurs at $T_{M2 \to M1}$. Histograms of the reconfiguration temperatures on repeated cycling at 0.1 °C/minute are shown in Fig. 2c. For this device $T_{M1 \to R}$ and $T_{M1 \to M2}$ are narrowly peaked at 66.4 °C and 65.3 °C respectively; for other devices different values are found. This can be explained by superheating of M1, which varies between devices because the ease of nucleation of the high temperature phase (R or M2) depends on microscopic details.

In contrast, $T_{R \to M1}$ and $T_{M2 \to M1}$ are both peaked at the same temperature of 65.0 °C, indicated by the dotted line in Fig. 2c. In a number of nanobeams of different sizes, grown on different occasions, these two temperatures always lay in the narrow range between 64.9 and 65.2 °C; moreover, neither storage in air for six months nor heating to 200 °C for an hour changed them, indicating that effects of oxygen vacancies[25] and hydrogen doping[26] were minimal. This observation can be explained as follows. A small amount of M1 is often visible at the interface in M2+R coexistence, probably because it reduces the elastic energy. On cooling there is therefore no need for nucleation of M1, and reconfiguration occurs as soon as the triple point is reached. In fact, the dynamics of this process can sometimes be observed. Fig. 2d shows a sequence of images taken in less than a second during the reconfiguration of a nanobeam after bringing it slowly down to 65.0 °C in M2+R coexistence. A small pre-existing wedge of M1 at the M2+R interface rapidly expands to completely replace the R part of the nanobeam. All the above observations thus suggest that the triple-point is between 64.9 and 65.2 °C.

We now consider varying $L$ at fixed $T$. First, in coexistence between any pair of phases the interface position is linear in $L$, as shown in Fig. 2e. This follows from the fact that the interface moves so as to maintain $P_c$ at the phase equilibrium value. A length increase $\delta L$ causes an interface shift $\delta y_{M1R}$ which changes the natural length by $\delta L$ to keep the strain constant. This implies $\delta L = \alpha_{M1R} \delta y_{M1R}$, where $\alpha_{M1R} \equiv c_{M1}/c_R - 1$. Hence $\delta y_{M1R}$ should vary according to $dL/dy_{M1R} = \alpha_{M1R}$, and likewise $dL/dy_{M2M1} = \alpha_{M2M1} \equiv c_{M2}/c_{M1} - 1$ and $dL/dy_{M2R} = \alpha_{M2R} \equiv c_{M2}/c_R - 1 \approx \alpha_{M2M1} + \alpha_{M1R}$. Best linear fits to the data shown give $\alpha_{M2M1} = 0.0074$, $\alpha_{M1R} = 0.0100$ and $\alpha_{M2R} = 0.0174$, close to the values of 0.0073, 0.0098, and 0.0172 calculated from the known lattice constants[28,29].

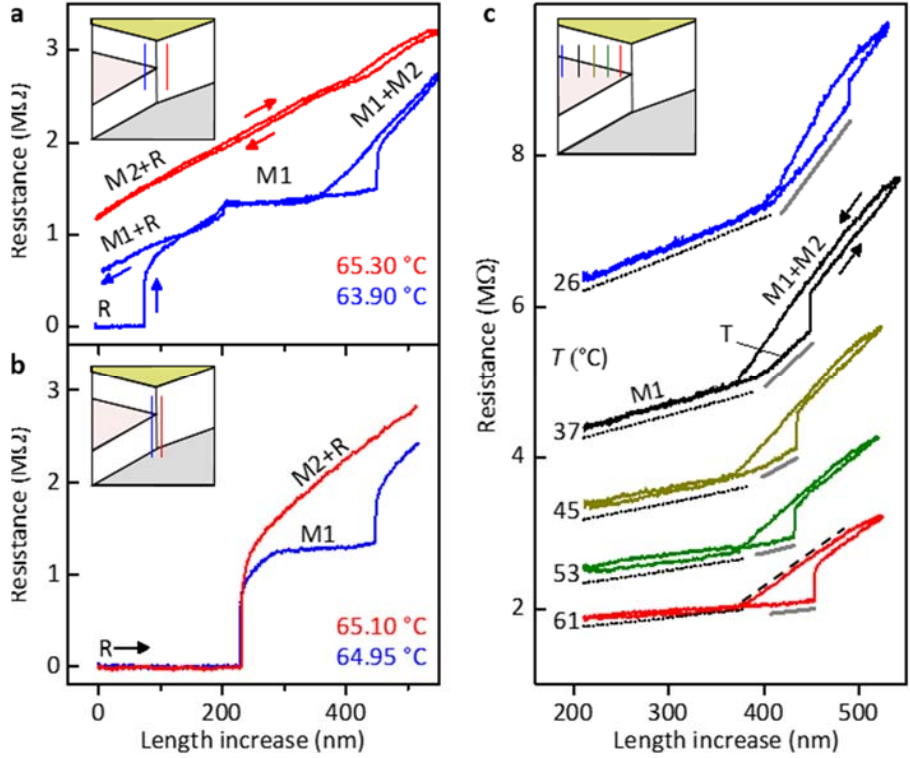

**Figure 3 | Resistance vs length measurements. a,** At 65.3 °C, above the triple point (red line in inset), the resistance varies steadily in M2+R coexistence. At 63.9 °C, below the triple point (blue line in inset), it reflects a sequence of transitions. (Device P10; $L$ is varied at 8 nm/s). **b,** Starting in the R state, M2 nucleates if $T \geq 65.10$ °C (red) whereas M1 nucleates if $T \leq 64.95$ °C (blue), implying that these lie on either side of $T_{tr}$ (see inset). **c,** The variation of the resistance with $L$ and $T$ is due to a strain-dependent activation energy in the M1 phase (dotted lines, offset by -0.15 MΩ for clarity) and to conversion of M1 to M2 in coexistence (dashed line). Gray lines indicate an additional resistance rise attributed to the T phase.

The ability to control $L$ allows us to confirm the temperature of the triple point and to determine the behavior very close to it. We exploit the fact that the electrical resistance $R_n$ is sensitive to the phase composition, since each phase has a different resistivity[12,13]. The measurements in Fig. 3 are for a device (P10) with indium contacts. Fig. 3a shows that at 65.3 °C $R_n$ changes smoothly with $L$, due to a smoothly changing M2+R interface position for $T > T_{tr}$ (see inset, red line). In contrast, at 63.9 °C it changes in a more complicated way, reflecting the sequence M1+R→M1→M2+M1 expected for $T < T_{tr}$ (see inset, blue line). Jumps and hysteresis here show that M1 and M2 both require nucleation, consistent with the transitions being first-order. To establish $T_{tr}$ we measured $R_n$ at a series of closely spaced temperatures, each time preparing the nanobeam in a fully metallic R state by cooling at sufficiently small $L$ for R to be stabilized by compression, and then increasing $L$ until an insulating domain nucleated. At 64.95 °C and below, the domain that appeared was always M1, while at 65.10 °C and above it was always M2, implying that $T_{tr}$ was between these two values (see Fig. 3b). This is perfectly consistent with the range of $T_{tr}$ deduced above from the $T$-sweeping measurements. Including uncertainties from variation between samples, temperature fluctuations and calibration, we conclude that $T_{tr} = 65.0 \pm 0.1$ °C.

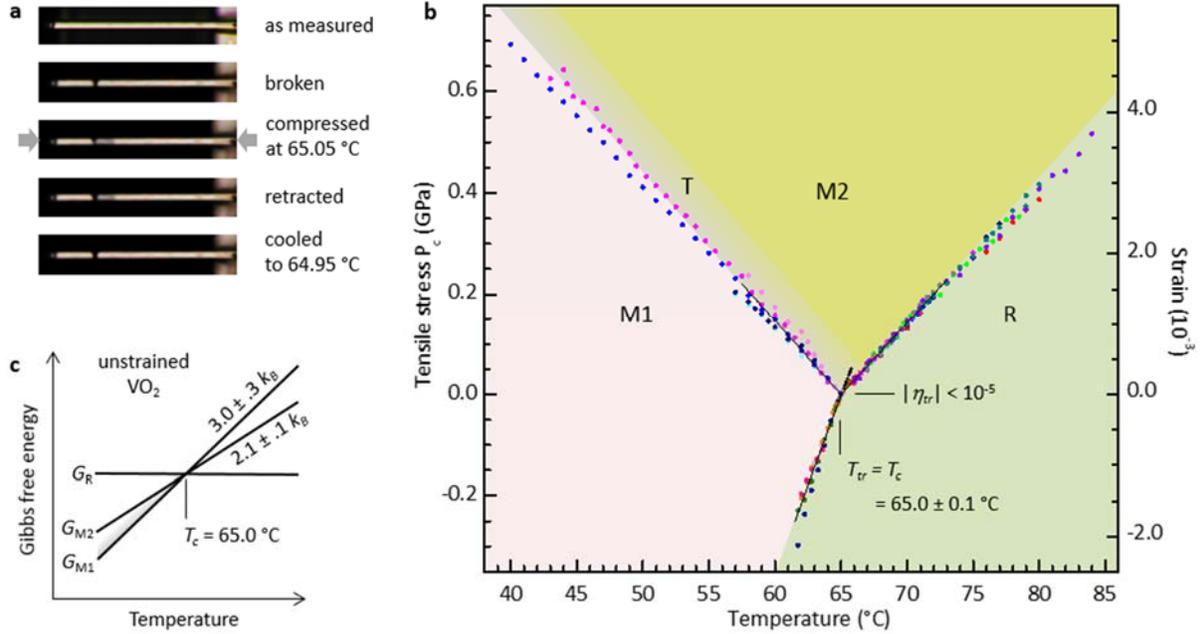

**Figure 4 | Phase diagram of VO$_2$. a,** The transition temperature $T_c$ at zero stress is measured by finding the temperature above which the metallic phase (darker) becomes stable in a cantilever, as illustrated here (device P8). It is found to be equal to the triple point temperature: $T_c = T_{tr} = 65.0 \pm 0.1$ °C. **b,** Deduced stress-temperature phase diagram. The small black filled circles are for superheated M1 phase. The gray shaded strip is where a metastable T phase can occur. **c,** The results imply that the free energies of all the phases are degenerate at $T_c$ in unstrained pure VO$_2$.

Resistance vs length measurements also yield other useful information, as illustrated in Fig. 3c (see Supplementary Information for details). First, the variation of the resistance of the M1 state with $L$ and $T$ is explained by a linear increase in the activation energy of the resistance with tensile strain $\eta = (L - L_0)/L_0$, $L_0$ being the effective natural length. The dotted lines are plots of $R_n \propto \exp[-(\Delta_0 + \gamma\eta)/k_B T]$ using coefficient values $\Delta_0 = 0.31$ eV and $\gamma = 0.77$ eV (the uncertainty in $\gamma$ is 10%), where $k_B$ is Boltzmann's constant. Second, from the variation of $R_n$ in M1+M2 coexistence (such as indicated by the dashed line) we can deduce that $\rho_{M2}/\rho_{M1} = 2.3 \pm 0.2$ and that the activation energies of M1 and M2 are the same to within a few percent. Third, a distinct additional rise in $R_n$, indicated by the solid gray lines, precedes the nucleation of M2 from M1. This can be explained by a continuous distortion of M1 into the T phase, which we immediately infer has a higher resistivity than M1 and is unstable relative to M2 at all temperatures from $T_{tr}$ to below 26 °C.

Although we cannot measure the axial stress $P_c$ directly, we can realize the condition $P_c = 0$ simply by breaking a nanobeam using a micromanipulator after other measurements have been completed. This produces opposing cantilevers, as illustrated in Fig. 4a. If the cantilevers are prepared in the fully M1 state by warming from lower temperature to around $T_c$ and are then brought together, the compression produces a domain of R phase in one of them. After retraction, this domain persists only above a certain temperature, and shrinks and disappears below it. We identify this temperature with $T_c$, the transition temperature at zero stress. By carrying out the procedure on a number of devices we obtained the striking result that in every nanobeam $T_c$ was equal to $T_{tr}$, to within an uncertainty of $\delta T \approx 0.05$ °C governed by temperature fluctuations. We thus conclude that $T_c = T_{tr} = 65.0 \pm 0.1$ °C.

Fig. 4b shows the phase diagram of VO$_2$ inferred from measurements on ten nanobeams (see Supplementary Information for details). Briefly, the $P_c(T)|_{ij}$ were deduced from measurements of $y_{ij}$ ($i,j$ =M1,M2,R) vs $T$ as follows. Since the stress in coexistence must take the phase equilibrium value, consideration of the variation of the strain $\eta = P_c/E$ with $T$ ($E$ is the Young's modulus, taken to be 140 GPa for every phase[6]) yields[13]

$$\frac{1}{E}\frac{\partial P_c}{\partial T}\bigg|_{ij} = \frac{d\eta}{dT}\bigg|_{ij} = -\frac{\alpha_{ij}}{L_0}\frac{dy_{ij}}{dT} - \Delta K \ . \tag{1}$$

The first term on the right is the change due to movement of the interface. The second, $\Delta K$, is the thermal expansion mismatch between nanobeam and silicon substrate, which produces a correction of 5-10 %. Given that $\eta(T_{tr}) = 0$, Eq. (1) can be used to derive $\eta(T)$ for each boundary. The deduced phase boundaries are straight, with uncertainties in their slopes of 5-10%, and obey the constraint at the triple point

$$\alpha_{M2M1}\frac{d\eta}{dT}\bigg|_{M2M1} + \alpha_{M1R}\frac{d\eta}{dT}\bigg|_{M1R} = \alpha_{M2R}\frac{d\eta}{dT}\bigg|_{M2R}, \tag{2}$$

which is imposed by the Clausius-Clapeyron relations,

$$\frac{\partial P_c}{\partial T}\bigg|_{ij} = \frac{S_j - S_i}{b^2(a_i - a_j)} \approx \frac{S_j - S_i}{\alpha_{ij}V}, \tag{3}$$

in combination with Eq. 1. Here $S_i$ is the entropy per vanadium pair in phase $i$, $b = 4.55$ Å is the base length of the rutile unit cell, and $V = 59$ Å$^3$ is the rutile unit cell volume. The value of $\partial P_c/\partial T|_{M1R} = 71$ MPa °C$^{-1}$ corresponds to the known latent heat[30] of 1020 cal/(mole formula unit); $\partial P_c/\partial T|_{M2R} = 29$ MPa °C$^{-1}$ corresponds to 710 cal/mole; and $\partial P_c/\partial T|_{M2M1} = -29$ MPa °C$^{-1}$. From the results we deduce entropy differences $S_R - S_{M1} = 3.0 \pm 0.3\ k_B$ and $S_R - S_{M2} = 2.1 \pm 0.1\ k_B$. The equality of $T_{tr}$ and $T_c$ to within $\delta T \approx 0.05$ °C implies that the strain $\eta_{tr}$ at the triple point is smaller than $\delta T \times d\eta/dT|_{M2R} = 1.0 \times 10^{-5}$, where $d\eta/dT|_{M2R} = 2.0 \times 10^{-4}$ °C$^{-1}$, and this is also indicated on the phase diagram. Finally, the finding that the T phase is metastable with respect to M2 is indicated by a gray shaded strip within the M2 stability region.

To stress the implication of these results we sketch in Fig. 4c the $T$ dependence of the Gibbs free energies $G_i$ of the phases of unstrained VO$_2$, setting $G_R = 0$. The slopes are the entropies $S_i = -dG_i/dT$ at zero stress. Precisely at the MIT the insulating M1 and M2 phases are simultaneously degenerate with the metallic R phase. This and other facts revealed by our measurements are not explained by current models of the transition, but will be crucial ingredients of the correct theory. For example, further development and application of the Landau theory[10] of VO$_2$ should be prompted by our results. The striking new insights we have gained into this important solid-state phase transition will be critical for both understanding and mastering the MIT in VO$_2$.

**Methods Summary**

VO$_2$ nanobeams grown by physical vapor transport were transferred onto slots on the micromachined silicon chips using a micromanipulator and bonded with UV-curable epoxy (Supplementary Information S1). Measurements from ten devices were used, and the temperature was calibrated using the known melting points of gallium and potassium (Supplementary S2). The slot width $L$ (20 or 40 μm) was varied piezoelectrically on a temperature stage under an optical microscope (Supplementary S3).

**Acknowledgments**

We thank Boris Spivak, Arkady Levanyuk, and Jiang Wei for helpful discussions. This work was supported by the U.S. Department of Energy, Office of Basic Energy Sciences, Division of Materials Sciences and Engineering, Award DE-SC0002197. The silicon chips were patterned at the UW Microfabrication Facility and the UCSB Nanofabrication Facility.

Supplementary Information for

# Measurement of a solid-state triple point at the metal-insulator transition in VO$_2$

Jae H. Park[1], Jim M. Coy[1], T. Serkan Kasirga[1], Chunming Huang[1], Zaiyao Fei[1], Scott Hunter[1], and David H. Cobden[1*]

[1]*Department of Physics, University of Washington, Seattle WA 98195, USA*

*Corresponding author: cobden@uw.edu

**S1. Device Fabrication**

Optical lithography and deep reactive-ion etching were used to micromachine slots in a 1 cm$^2$ silicon chip with 2 µm wet oxide to form the paddle. VO$_2$ nanobeams, typically 50-100 µm long and <0.5 µm thick, grown by physical vapor transport[1] on SiO$_2$ from a V$_2$O$_5$ source and loosened with buffered oxide etch, were placed across the slot using a micromanipulator. UV-curable epoxy was then applied to the both ends of the nanobeam and cured. The epoxy wets the interface between nanobeam and chip, effectively clamping the nanobeam close to the edges of the slot. In some cases the nanobeam was contacted electrically by drawing molten indium wires, gold-wire-bonded to Au/Ti contacts which were patterned using shadow-mask evaporation (as in Fig. 1c). In one case electron-beam induced Pt deposition was used both to clamp the nanobeam and to make contacts, but the Pt pads quickly fractured.

**S2. Temperature calibration and reproducibility**

Several PT100 platinum sensors were configured for four-terminal sensing and calibrated against the melting and boiling points of water. Their readings were found to be accurate to within 0.1 °C. One sensor was embedded in the invar stage. The stability of the measured temperature was better than ±0.05 °C. To correct for a temperature differential between the sample and sensor, due mainly to heat loss to the surrounding air, we placed small oil-coated crystals of the metals gallium (melting point 29.77 °C) and potassium (63.38 °C) on a silicon chip close to a nanobeam. Their melting could be seen under the microscope, and the calibration of the system was adjusted accordingly. This procedure was carried out on several chips. In this way the temperature difference between the chip body and the paddle was found to be less than 0.1 °C. Since the transition temperature that we found was on the lower side of the range of values in the literature, we carried out a number of checks. In the most rigorous, we constructed a separate aluminium thermal stage, with a different sensor, in an air-tight chamber having a glass lid through which the melting and the MIT in an unstrained device could be seen. The results were all consistent with $T_c = 65.0 \pm 0.1$ °C.

The reproducibility of the temperature measurements is illustrated in Table S1, in which we show the peak positions in the histograms of the jump temperatures (as illustrated in Fig. 2c) for six devices. Sample images of the same devices are also shown. The images were taken after extensive measurement runs during which in some cases (noteably P4B and P14) surface contamination had accumulated. This contamination had no discernable effect on the MIT behavior.

| Device | $T_{R \to M1}$ (°C) | $T_{M1 \to R}$ (°C) | |
|---|---|---|---|
| P4B | 64.92 | 67.44 | 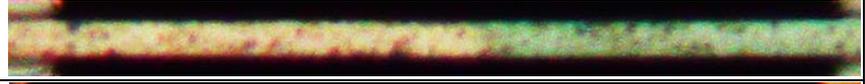 |
| P7 | 65.20 | 65.70 | 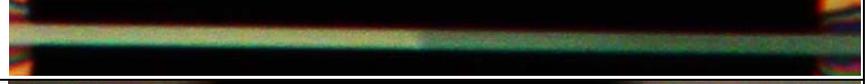 |
| P8B | 64.99 | 65.20 | 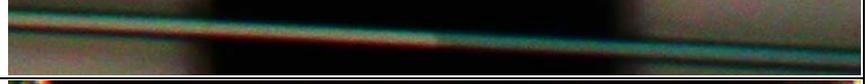 |
| P10C | 65.07 | 67.46 | 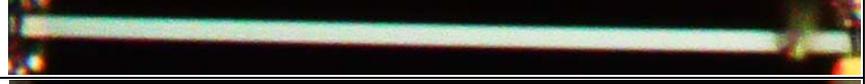 |
| P12 | 64.93 | 65.32 | 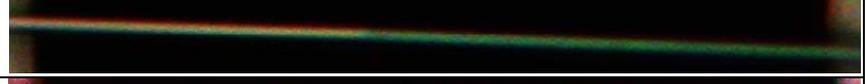 |
| P14 | 64.98 | 65.26 | 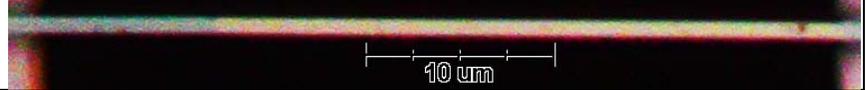 |

Table S1. Jump temperatures and images in metal-insulator coexistence of six of the devices on which detailed measurements were made. P8b is across a 20 μm slot; the others are across 40 μm slots.

## S3. Strain application and measurement

The home-built temperature-strain setup contains an invar (to minimize thermal expansion) mounting block under a high-power optical microscope in air. A heater and sensor are separately embedded in the block and the temperature is actively controlled. After the chip is positioned and held down with a clip, the slot width is controlled by displacing the paddle which rotates around the thin hinge stalk. The paddle is pushed using a single-axis horizontal piezo-actuator and the slot width can be increased or decreased by pushing on either one or other side of the paddle (the latter case is shown in Fig. 1c). A collimated and intensity-modulated laser diode beam is reflected from the side wall of the paddle and its deflection measured using a position sensitive detector (PSD), to detect the angular deflection of the paddle. The PSD signal modulation is calibrated to the slot width by recording the full-range deflection of the paddle which can be seen in the microscope. The setup could also be retrospectively calibrated using measurements such as those in Fig. 2e of the coexistence interface position combined with the known lattice constants of the phases.

## S4. Constructing the phase diagram

The literature values of the chain-axis lattice constants[2] are $c_R = 5.700$ Å (double the tetragonal c-axis lattice constant in rutile, measured at 66 °C), $c_{M1} = 5.755$ Å (the monoclinic a-axis lattice constant in M1, at 65 °C), and $c_{M2} = 5.797$ Å (the monoclinic b-axis lattice constant in M2, stabilized by Cr doping, at 25 °C). From these we obtain $\alpha_{M1R} \equiv c_{M1}/c_R - 1 = 0.0098$, $\alpha_{M2R} \equiv c_{M2}/c_R - 1 = 0.0172$ and $\alpha_{M2M1} \equiv c_{M2}/c_{M1} - 1 = 0.0073$, in good agreement with the values deduced from interface motion measurements such as those shown in Fig. 2e. Since the $\alpha$'s are small we have, to a percent accuracy, $\alpha_{M2R} = \alpha_{M2M1} + \alpha_{M1R}$. The shape of the phase boundary between phases $i$ and $j$ ($i,j$ =M1,M2,R) is deduced from measurements of interface position $y_{ij}$ vs $T$ as follows. The stress in coexistence should be nearly uniform, thanks to the geometry of a straight thin beam, and therefore equal to the phase equilibrium value at the interface, $P_c(T)$. If the Young's modulus $E$ is the same in each phase then the axial strain has a uniform value $\eta = P_c/E$ throughout the beam and varies with temperature according to[3]

$$\frac{1}{E}\frac{\partial P_c}{\partial T}\bigg|_{ij} = \frac{d\eta}{dT}\bigg|_{ij} = -\frac{\alpha_{ij}}{L_0}\frac{dy_{ij}}{dT} - \Delta K . \quad (1)$$

The first term on the right is the change in strain due to movement of the interface, $L_0$ being the effective clamped length.

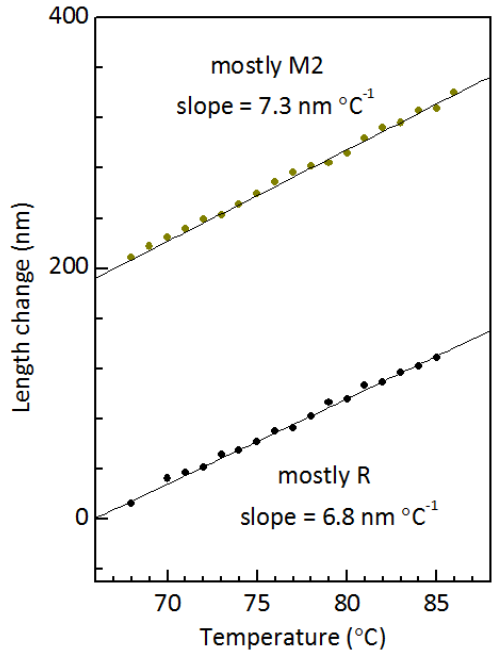

Figure S1. Measurements of length change needed to maintain a fixed M2-R interface position for device P14, yielding a difference in thermal expansion coefficients $K_R - K_{M2} \approx 0.17 \times 10^{-4}$ °C$^{-1}$ (see text).

The second term, $\Delta K = K_{VO_2} - K_{Si}$, is the thermal expansion mismatch between the nanobeam and the silicon substrate, where $K_{Si} = 0.03 \times 10^{-4}$ °C$^{-1}$. (The slot expands with temperature along with monolithic silicon chip that it is etched in.) For the monoclinic phases[4], $K_{VO_2} \approx 0.12 \times 10^{-4}$ °C$^{-1}$ giving $\Delta K \approx 0.09 \times 10^{-4}$ °C$^{-1}$. For the R phase[4], $K_{VO_2} \approx 0.30 \times 10^{-4}$ °C$^{-1}$, but we typically perform measurements with about half the nanobeam metallic, in which case $\Delta K \approx 0.2 \times 10^{-4}$ °C$^{-1}$. Hence $\Delta K$ gives a correction to the phase boundary slope of less than 10 %. As the interface moves and the fraction of R phase increases this correction increases and its magnitude can account for the slight downward curvature of the phase boundary measurements plotted in Fig. 4b. We chose not to attempt to compensate for the effect because the thermal expansion coefficients are not known accurately and the effect is not larger than the spread in the measurements between devices. In fact, we can determine the difference in $K_{VO_2}$ between phases $i$ and $j$ by comparing measurements made for the nanobeam mainly in phase $i$ with those for it mainly in phase $j$. For example, Fig. S1 shows measurements in M2-R coexistence on device P14 ($L_0 = 40$ μm) in which $T$ was varied while simultaneously varying $L$ so as to keep $y_{M2R}$ constant, with (i) the suspended section mostly R, and (ii) with it 75% M2 (ie, with the interface close to either one contact or the other). The difference between the slopes obtained, about 0.5 nm °C$^{-1}$, according to Eq. 1 results solely from the difference between $K_R$ and $K_{M2}$, and should be about $0.75 L_0 \times (K_R - K_{M2})$. From this we get $(K_R - K_{M2}) \approx (0.5$ nm °C$^{-1})/(30$ μm$) \approx 0.17 \times 10^{-4}$ °C$^{-1}$, in good agreement with the literature values given above.

We believe the assumption made in Eq. 1 of a single Young's modulus for the three all phases is justified because the nature of the bonding in the material does not differ much between the phases, and because it leads to consistent results. A recent experiment[5] based on measurements of a nanobeam reported possible evidence of a difference between M1 and M2 at the 10-20% level, which would not significantly affect any of our results. Moreover, even if $E$ were substantially different between phases it would make the equations more complicated and could alter the phase boundary slopes somewhat but would have no effect on our main conclusions.

To obtain $\eta(T)$ for each boundary from the $y_{ij}$ measurements using Eq. 1 we first subtracted a constant from each $y_{ij}(T)$ dataset so as to set $y_{ij} = 0$ at $T_{tr} = 65.0$ °C. We then divided the results by the gap width $L_0$ which was close to either 20 or 40 μm (some allowance was made for a slightly different apparent clamping length). Only measurements on devices with a clear single moving interface were used. The deduced boundary slopes at the triple point, shown by the solid straight lines in Fig. 4b, were constrained to be consistent with

$$\alpha_{M2M1} \left.\frac{\partial \eta}{\partial T}\right|_{M2M1} + \alpha_{M1R} \left.\frac{\partial \eta}{\partial T}\right|_{M1R} = \alpha_{M2R} \left.\frac{\partial \eta}{\partial T}\right|_{M2R} . \quad (2)$$

This results from the relevant Clausius-Clapeyron relations,

$$\left.\frac{\partial P_c}{\partial T}\right|_{ij} = \frac{S_j - S_i}{b^2(a_i - a_j)} \approx \frac{S_j - S_i}{\alpha_{ij} V}, \quad (3)$$

where $S_i$ is the entropy per vanadium pair. Here $b = 4.55$ Å is the base length and $V = b^2 a_R = 59$ Å$^3$ is the volume of the tetragonal unit cell. Eliminating the entropies, using Eq. 1 taking $\Delta K$ and $E$ to be constants, and using $\alpha_{M2R} \approx \alpha_{M2M1} + \alpha_{M1R}$ gives Eq. 2. The stress $P_c$ was then obtained by multiplying by $E = 140$ GPa. The straight line fit to the M1-R phase boundary shown in Fig. 4b has slope $\partial P_c/\partial T|_{M1R} = 71$ MPa °C$^{-1}$ which corresponds to a specific latent heat $T_c(S_R - S_{M1})/V = 1020$ cal/(mole formula unit) that is the same as measured previously[6] for the MIT in a macroscopic crystal.

## S5. Electrical measurements

Here we elaborate on the analysis of the measurements of resistance $R_n$ vs length $L$ and temperature $T$ in Figure 3. The electrical contacts to device P10 were made by drawing molten indium at the edges of the slot using a nanomanipulator, and there was a substantial contact resistance $R_c$, probably because the indium partially oxidized. $R_c$ was determined to be 0.9 MΩ from the value of the resistance when the nanobeam was in the fully metallic R state (since the resistivity of R is $10^4$ times smaller than M1 or M2, $\rho_R \approx 3 \times 10^{-4}$ Ω cm) and was subtracted from $R_n$.

The conductivities of M1 and M2 are known to be activated, though their values vary somewhat in the literature. In earlier work[3] we found the resistivity of M2 to have a well defined value of $\rho_{M2} \approx 12$ Ωcm at the MIT, and its activation energy in unstrained (buckled) nanobeams to be $\Delta_{M2} = 0.30 \pm 0.01$ eV. The variation of $R_n$ in the M1 state in Fig. 3c is consistent with a linear variation of the activation energy with strain, i.e.:

$$R_n = R_{M1} = R_0 \exp(\Delta_{M1}/kT),$$

with
$$\Delta_{M1} = \Delta_0 + \gamma\eta.$$

Here $\eta = (L - L_0)/L_0$, where $L_0$ is the natural effective clamped length, and $\Delta_0$ is the gap at $\eta = 0$. We determine the slot width $L$ at $\eta = 0$ using our knowledge that the triple point is at zero strain (it is at length increase ≈ 300 nm in Fig. 3). We obtain the best fit to the entire dataset, shown by the dotted lines in Fig. 3c (which are offset by -0.15 MΩ for clarity), with parameter values $\Delta_0 = 0.31$ eV, $\gamma = 0.77$ eV, and $R_0 = 40.5$ MΩ. The uncertainty is about 0.01 eV in $\Delta_0$ and 10% in $\gamma$. From optical images we roughly estimate the cross-sectional area of the nanobeam as $A \approx 1$ μm², putting $\rho_{M1} = R_n A/L_0$ in the ballpark of (2 MΩ)(1 μm²)/(40 μm) ~ 5 Ωcm.

The variation of $R_n$ in the M1+M2 state in Fig. 3c is due to the change in the interface position $y_{M2M1}$ from 0 to $L_0$ as the nanobeam gradually converts from M1 to M2. We can deduce the ratio of their resistivities using

$$R_n = R_{M1} + (R_{M2} - R_{M1})\frac{y_{M2M1}}{L_0},$$

from which we have

$$\frac{1}{R_{M1}}\frac{dR_n}{dL} = \left(\frac{R_{M2}}{R_{M1}} - 1\right)\frac{1}{L_0}\frac{dy_{M2M1}}{dL} = \left(\frac{\rho_{M2}}{\rho_{M1}} - 1\right)\frac{1}{L_0}\frac{1}{\alpha_{M2M1}}$$

and thus

$$\frac{\rho_{M2}}{\rho_{M1}} = 1 + \frac{\alpha_{M2M1}L_0}{R_{M1}}\frac{dR_n}{dL}.$$

For example, using the data at 61 °C we have $\frac{dR_n}{dL} = 8.6$ MΩ/μm (the slope of the dashed line in Fig. 3c), $R_{M1} = 2.0$ MΩ (taken from $R_n$ at the foot of the dashed line where $y_{M2M1} = 0$), $\alpha_{M2M1} = 0.0074$, and $L_0 = 40$ μm, which gives $\rho_{M2}/\rho_{M1} = 2.3$ with about 10% uncertainty. This ratio has not been determined accurately before. This again gives $\rho_{M1} = (12\ \Omega\text{cm})/2.3 \approx 5$ Ωcm. In addition, we find that $\rho_{M2}/\rho_{M1}$ does not change by more than 5% between 26 and 64 °C, implying that the activation energies are equal, $\Delta_{M2} = \Delta_{M1}$, to within a few percent.